\documentclass[amsmath,amssymb,showpacs]{revtex4}

\usepackage{graphicx}
\usepackage{dcolumn}
\usepackage{bm}
\newcommand{\be}{\begin{equation}}
\newcommand{\ee}{\end{equation}}
\newcommand{\ba}{\begin{eqnarray}}
\newcommand{\ea}{\end{eqnarray}}
\newcommand{\baa}{\begin{eqnarray*}}
\newcommand{\eaa}{\end{eqnarray*}}

\newcommand{\dis}{\displaystyle}

\begin{document}

\title{
Replica-Exchange Method in van der Waals Radius Space: \\
Overcoming Steric Restrictions for Biomolecules
} 

\author{
Satoru G. Itoh\footnote{Electronic address: itoh@tb.phys.nagoya-u.ac.jp},
Hisashi Okumura\footnote{Current Address:
Research Center for Computational Science, Institute for Molecular Science,
Okazaki, Aichi 444-8585, Japan, Electronic address: hokumura@ims.ac.jp}, 
and Yuko Okamoto\footnote{Electronic address: okamoto@phys.nagoya-u.ac.jp}
}
\affiliation{
Department of Physics \\
School of Science \\
Nagoya University \\ 
Nagoya, Aichi 464-8602, Japan \\
}

\begin{abstract}

We present a new type of the Hamiltonian replica-exchange method, 
in which not temperatures but the van der Waals radius parameter
is exchanged. 
By decreasing the van der Waals radii that control spatial 
sizes of atoms, 
this Hamiltonian replica-exchange method overcomes 
the steric restrictions and energy barriers.
Furthermore, the simulation based on this method escapes from 
the local-minimum free-energy states and realizes effective sampling 
in the conformational space. 
We applied this method to an alanine dipeptide in aqueous solution and 
showed the effectiveness of the method by comparing the results 
with those obtained from the conventional canonical method. 

\end{abstract}

\maketitle

%

  \section{Introduction} 
  \label{intro:sec}

Effective samplings in the conformational space 
by Monte Carlo (MC) and molecular dynamics (MD)
simulations
are necessary to predict the native structures of proteins.
In the conventional canonical-ensemble 
simulations \cite{mrrtt53,hlm82,evans83,nose_mp84,nose_jcp84,hoover85}, 
however, 
it is difficult to realize effective samplings in complex systems 
such as proteins. 
This is because the usual canonical-ensemble simulations tend to get 
trapped in a few of many local-minimum states. 
To overcome these difficulties, various generalized-ensemble algorithms 
have been proposed (for reviews, see, e.g., Refs.~\cite{mso01,ioo07}). 

The replica-exchange method (REM) \cite{hn96} 
is one of the most well-known methods among the generalized-ensemble 
algorithms (see Ref.~\cite{so99} for the MD version). 
It is easier to implement than the multicanonical 
algorithm \cite{berg91,berg92}, 
which is also one of the most well-known generalized-ensemble algorithms
(see Refs.~\cite{hoe96,naka97} for the MD version), 
because we do not have to determine a probability weight factor 
in advance in the REM. 
In the multicanonical and 
similar algorithms 
\cite{oo04a,oo04b,oo04c,oo04d,oo06,berg03,itoh04,itoh06,itoh07a,itoh07b} 
we employ non-Boltzmann weight factors as the probability weight factors. 
These non-Boltzmann weight factors are not {\it a priori} known and 
have to be determined by tedious procedures. 
On the other hand, 
the usual Boltzmann weight factor is employed in REM, and therefore 
it is not necessary to determine the non-Boltzmann weight factor. 
The REM uses non-interacting replicas of the target system with 
different temperatures and realizes a random walk in temperature space 
by exchanging the temperatures of pairs of replicas. 
Accordingly, the simulation can avoid getting trapped in 
local-minimum free-energy states. 

For large systems such as proteins in aqueous solution, however, 
the usual REM has a difficulty. 
We need to increase the number of replicas in proportion 
to $O(f^{\frac{1}{2}})$, 
where $f$ is the number of degrees of freedom \cite{hn96}. 
Large biomolecular systems, therefore, require a large number of
replicas in the REM and hence huge amount of computation time.
In order to overcome this difficulty, 
it was pointed out that the number of required replicas
can be greatly decreased if only the parameter exchanges
are performed in the multi-dimensional replica-exchange
method (MREM) \cite{sko00} without temperature exchanges
\cite{fwt02}.
In MREM replica exchanges in temperature and/or parameter
in the potential energy are performed.
MREM is also referred to as the Hamiltonian replica-exchange
method \cite{fwt02}, and in the present article we use the
latter terminology.
   
When we perform simulations of a protein in explicit water solvent, 
most of the degrees of freedom is occupied by water molecules.
In order to predict the native structure of a protein, for instance, 
we would like to sample effectively the conformational space of the protein 
rather than water molecules. 
As an application of the Hamiltonian REM, therefore, 
Berne and co-workers performed simulations of the peptide 
in explicit water solvent, 
in which the scales of the only interactions related to 
the protein are varied \cite{lkfb05}. 
They could achieve effective samplings in the conformational space 
of the peptide and 
saved CPU cost in comparison with the usual REM. 
Moreover, another application of the Hamiltonian REM was reported 
by Kannan and Zacharias \cite{kz07}. 
They focused on the backbone dihedral angles of the peptides and 
added biasing potential energy to the backbone dihedral angles. 
They could also achieve effective samplings in this space with 
the biasing potential energy. 
However, these biasing potential energy terms are complicated 
functions and highly dependent on force fields. 

In this article, we propose a new type of Hamiltonian REM 
where we exchange the scaling factor of the van der Waals
radius of solute atoms in the interaction terms among
solute atoms only.  By reducing this scaling factor, the
steric hindrance among solute atoms will be reduced and
wide conformational space can be explored.
We applied this method to the sytems of an alanine dipeptide 
with explicit water molecules and 
tested the effectiveness of the method by comparing the results 
with those from the conventional canonical method. 

In Section \ref{method:sec} we describe the new Hamiltonian REM. 
We give the details of the new Hamiltonian REM 
and canonical simulations 
that we performed in Section \ref{det:sec}. 
The results are presented in Section \ref{comp:sec}. 
Section \ref{conc:sec} is devoted to conclusions. 

  \section{Methods} 
  \label{method:sec}

  \subsection{Hamiltonian REM} 
  \label{hrem:subsec}
We first give a general formulation of the
Hamiltonian REM \cite{sko00}.

We consider a system of {\it N} atoms with their coordinate vectors and 
momentum vectors denoted by 
$q \equiv \left\{ {\mbox {\boldmath $q$}}_{1},\cdots,{\mbox {\boldmath $q$}}_{N} \right\}$ and 
$p \equiv \left\{ {\mbox {\boldmath $p$}}_{1},\cdots,{\mbox {\boldmath $p$}}_{N} \right\}$, respectively. 
The Hamiltonian $H_{\lambda}$ in state $x \equiv (q,p)$ is given by
the sum of the kinetic 
energy {\it K} and 
potential energy $E_{\lambda}$: 
\begin{eqnarray} 
H_{\lambda}(x) = K(p)+E_{\lambda}(q)~.
\label{hamiltonian}
\end{eqnarray}
Here, we are explicitly writing (or introducing) a parameter of interest
in the potential energy as $\lambda$.
In the canonical ensemble at temperature $T$,  
each state $x$ 
is weighted by the Boltzmann factor:
\begin{equation}
W_{\rm B}(x) = e^{-\beta H_{\lambda}(x)}~,
\label{eqn3}
\end{equation}
where the inverse temperature $\beta$ is defined by 
$\beta = 1/k_{\rm B} T$ ($k_{\rm B}$ is Boltzmann's constant). 

The generalized ensemble for the Hamiltonial REM consists of 
$M$ {\it non-interacting} copies (or, replicas) 
of the original system in the canonical ensemble
at $M$ different parameter values $\lambda_m$ ($m=1, \cdots, M$).
We arrange the replicas so that there is always
exactly one replica at each $\lambda$ value.
Then there is a one-to-one correspondence between replicas
and parameter values; the label $i$ ($i=1, \cdots, M$) for replicas 
is a permutation of 
the label $m$ ($m=1, \cdots, M$) for $\lambda_m$,
and vice versa:
\begin{equation}
\left\{
\begin{array}{rl}
i &=~ i(m) ~\equiv~ f(m)~, \cr
m &=~ m(i) ~\equiv~ f^{-1}(i)~,
\end{array}
\right.
\label{eqn4b}
\end{equation}
where $f(m)$ is a permutation function of $m$ and
$f^{-1}(i)$ is its inverse.

Let $X = \left\{x_1^{[i(1)]}, \cdots, x_M^{[i(M)]}\right\} 
= \left\{x_{m(1)}^{[1]}, \cdots, x_{m(M)}^{[M]}\right\}$ 
stand for a ``state'' in this generalized ensemble.
Here, the superscript $i$ and the subscript $m$ in $x_m^{[i]}$
label the replica and the parameter, respectively.
The state $X$ is specified by the $M$ sets of 
coordinates $q^{[i]}$ 
and momenta $p^{[i]}$
of $N$ atoms in replica $i$ at parameter $\lambda_m$:
\begin{equation}
x_m^{[i]} \equiv \left(q^{[i]},p^{[i]}\right)_m~.
\label{eqn5}
\end{equation}
Because the replicas are non-interacting, the weight factor for
the state $X$ in
this generalized ensemble is given by
the product of Boltzmann factors for each replica $i$ (or at each
parameter $\lambda_m$):
\begin{equation}
W_{\rm HREM}(X) = 
\dis{\prod_{i=1}^M} 
\dis{\exp \left\{ 
- \beta H_{\lambda_{m(i)}}\left( x_{m(i)}^{[i]} \right) \right\} }
 = \dis{\prod_{m=1}^M} 
\dis{\exp \left\{
- \beta H_{\lambda_{m}}\left( x_m^{[i(m)]} \right) \right\}~, }
\label{eqn7}
\end{equation}
where $i(m)$ and $m(i)$ are the permutation functions in 
Eq.~(\ref{eqn4b}).

We now consider exchanging a pair of replicas in the generalized
ensemble.  Suppose we exchange replicas $i$ and $j$ which are
at parameter values $\lambda_m$ and $\lambda_n$, respectively:  
\begin{equation}
X = \left\{\cdots, x_m^{[i]}, \cdots, x_n^{[j]}, \cdots \right\} 
\longrightarrow \ 
X^{\prime} = \left\{\cdots, x_m^{[j]}, \cdots, x_n^{[i]}, 
\cdots \right\}~. 
\label{eqn8}
\end{equation}
The transition probability for this replica exchange process is 
given by the usual Metropolis criterion: 
\begin{equation}
w(X \rightarrow X^{\prime})
={\rm min}\left(1,\frac{W_{\rm HREM}(X^{\prime})}
{W_{\rm HREM}(X)}\right)
={\rm min}\left(1,\exp(-\Delta)\right)~,
\label{eqn18}
\end{equation}
where we have \cite{sko00}
\begin{equation}
\Delta = \beta \left[ \left( E_{\lambda_m} \left(q^{[j]}\right) - 
E_{\lambda_m} \left(q^{[i]}\right) \right)
- \left( E_{\lambda_n} \left(q^{[j]}\right) - 
E_{\lambda_n} \left(q^{[i]}\right) \right) \right]~.
\label{eqn32}
\end{equation}
Here, $q^{[i]}$ and $q^{[j]}$ stand for coordinate vectors
for replicas $i$ and $j$, respectively, before the replica exchange.
Note that we need to newly evaluate the potential energy for
exchanged coordinates, $E_{\lambda_m}(q^{[j]})$ and
$E_{\lambda_n}(q^{[i]})$, because $E_{\lambda_m}$ and $E_{\lambda_n}$
are in general different functions.
We remark that the kinetic energy terms have canceled out each other
in Eq.~(\ref{eqn32}).
The Hamiltonian REM is realized by alternately performing the following 
two steps \cite{sko00}: 
\begin{enumerate}
\item  For each replica, a canonical MC or MD simulation at the 
corresponding
parameter value $\lambda_m$ is carried out simultaneously and independently 
for a certain steps
with the corresponding Boltzmann factor of Eq.~(\ref{eqn3}) for each replica. 
\item 
We exchange a pair of replicas $i$ and $j$ which are at the
parameter values $\lambda_m$ and $\lambda_n$, respectively.
The transition probability for this replica exchange process is given by 
Eqs.~(\ref{eqn18}) and (\ref{eqn32}).
\end{enumerate}

Finally, we remark that in order to further enhance sampling, we can
always introduce a temperature-exchange process together with
the above parameter exchange \cite{sko00}.

  \subsection{van der Waals Replica-Exchange Method} 
  \label{lj:subsec}

We now describe our special realization of the Hamiltonian REM,
which we refer to as the
{\it van der Waals Replica-Exchange Method (vWREM)}.
  
We consider a system consisting of solute molecule(s)
in explicit solvent. 
We can write the total potential energy as follows.
\begin{eqnarray} 
E_{\lambda}(q) = E_{\rm p}(q_{\rm p}) + E_{\rm ps}(q_{\rm p},q_{\rm s})
+ E_{\rm s}(q_{\rm s})~,
\label{potential}
\end{eqnarray}
where $E_{\rm p}$ is the potential energy for
the atoms in the solute only, $E_{\rm ps}$ is
the interaction term between solute atoms and
solvent atoms, and $E_{\rm s}$ is the potential
energy for the atoms of the solvent molecules only.
Here, $q=\{q_{\rm p},q_{\rm s}\}$, where
$q_{\rm p}$ and $q_{\rm s}$ are 
the coordinate vectors of the solute atoms and
the solvent atoms, respectively, and denoted by
$q_{\rm p} \equiv \left\{ {\mbox {\boldmath $q$}}_{1},\cdots,
{\mbox {\boldmath $q$}}_{N_{\rm p}} \right\}$ and
$q_{\rm s} \equiv \left\{ {\mbox {\boldmath $q$}}_{N_{\rm p}+1},\cdots,
{\mbox {\boldmath $q$}}_{N} \right\}$. 
($N_{\rm p}$ is the total number of atoms in the solute.)

We are more concerned with effective sampling
of the conformational space of the solute itself 
than that of the solvent molecules. 
The steric hindrance of the solute conformations are governed
by the van der Waals radii of each atom in the solute.
Namely, when the van der Waals radii are large, 
the solute molecule is bulky and we have more steric
hindrance among the solute atoms by the Lennard-Jones
interactions, and when it is small,
the solute molecule can move more freely.
We thus introduce a parameter $\lambda$ that scales the
van der Waals radius of each atom in the solute by
\begin{equation}
\sigma_{k \ell} \longrightarrow \lambda \sigma_{k \ell}~,
\label{scaleLJ}
\end{equation}
and write the Lennard-Jones energy term within $E_{\rm p}$
in Eq.~(\ref{potential}) as follows:
\begin{eqnarray}
V_{\lambda}\left(q_{\rm p}\right) 
= \sum^{N_{\rm p}-1}_{k=1}\sum^{N_{\rm p}}_{\ell=k+1} 
4{\epsilon_{k \ell}} 
\left\{\left(\frac{\lambda \sigma_{k \ell}}{r_{k \ell}} \right)^{12}- 
\left(\frac{\lambda \sigma_{k \ell}}{r_{k \ell}} \right)^{6} \right\}~, 
\label{pot_LJ_protein}
\end{eqnarray}
where $r_{k \ell}$ is the distance between atoms $k$ and $\ell$
in the solute 
and $\epsilon_{k \ell}$ and $\sigma_{k \ell}$ are the corresponding
Lennard-Jones parameters. 
The original potential energy is recovered when $\lambda = 1$, and
the steric hindrance of solute conformations is reduced when
$\lambda < 1$. 
Note that this is the only $\lambda$-dependent term in
$E_{\lambda}$ in Eq.~(\ref{potential}).

We prepare $M$ values of $\lambda$, $\lambda_m$ ($m=1, \cdots, M$).
Without loss of generality, we can assume that the parameter
values are ordered as $\lambda_1 < \cdots < \lambda_M$.
The vWREM is realized by alternately performing the following 
two steps: 
\begin{enumerate}
\item  For each replica, a canonical MC or MD simulation at the 
corresponding
parameter value $\lambda_m$ is carried out simultaneously and independently 
for a certain steps
with the corresponding Boltzmann factor of Eq.~(\ref{eqn3}) for each replica. 
\item 
We exchange a pair of replicas $i$ and $j$ which are at the
neighboring parameter values $\lambda_m$ and $\lambda_{m+1}$, respectively.
The transition probability for this replica exchange process is given by 
Eq.~(\ref{eqn18}), where $\Delta$ in Eq.~(\ref{eqn32}) now reads
\begin{equation}
\Delta = \beta \left[ \left( V_{\lambda_m}\left(q^{[j]}_{\rm p}\right) - 
V_{\lambda_m}\left(q^{[i]}_{\rm p}\right) \right)
- \left( V_{\lambda_{m+1}}\left(q^{[j]}_{\rm p}\right) - 
V_{\lambda_{m+1}}\left(q^{[i]}_{\rm p}\right) \right) \right]~.
\label{deltaVp}
\end{equation}
Here, $V_{\lambda}$ is the Lennard-Jones
potential energy in 
in Eq.~(\ref{pot_LJ_protein})
among the solute atoms only.
\end{enumerate}
Note that because the $\lambda$ dependence of $E_{\lambda}$
exists only in $V_{\lambda}$, the rest of the terms have
been cancelled out in Eq.~(\ref{eqn32}).
 
We see that Eq.~(\ref{deltaVp}) includes only the 
coordinates $q_{\rm p}$ of the atoms
in the solute only and is independent of the 
coordinates $q_{\rm s}$ of solvent molecules. 
Because $N_{\rm p} \ll N$ usually holds, 
the difficulty in the usual REM that 
the number of required replicas increases with 
the number of degrees of freedom is much alleviated in this formalism. 

  \subsection{Reweighting Techniques} 
  \label{reweight:subsec}

The results from Hamiltonian REM simulations with different parameter values 
can be analyzed by the reweighting techniques \cite{fs89,kbskr92}. 
Suppose that we have carried out a Hamiltonian REM simulation 
at a constant temperature $T_{0}$ with $M$ replicas 
corresponding to $M$ parameter values $\lambda_m$ ($m=1,\cdots,M$).
   
For appropriate reaction coordinates $\xi_{1}$ and $\xi_{2}$, 
the canonical probability distribution $P_{T,\lambda}(\xi_{1},\xi_{2})$ 
with any parameter value $\lambda$ at any 
temperature $T$ can be calculated from 
\begin{eqnarray}
P_{T,\lambda}(\xi_{1},\xi_{2}) = \sum_{E_{\lambda_1},\cdots,E_{\lambda_M}}
\frac{\displaystyle \sum^{M}_{m=1} \left( g_{m} \right)^{-1} 
N_{m}(E_{\lambda_1},\cdots,E_{\lambda_M};\xi_{1},\xi_{2})
e^{-\beta E_{\lambda}}}
{\displaystyle \sum^{M}_{m=1} \left( g_{m} \right)^{-1} 
n_{m} e^{f_{T_0,\lambda_m}-\beta_{0} E_{\lambda_m}}}~,
\label{prob}
\end{eqnarray}
and 
\begin{eqnarray}
e^{-f_{T_0,\lambda_m}} = \sum_{\xi_{1},\xi_{2}} P_{T_0,\lambda_m}(\xi_{1},\xi_{2})~.
\label{dless_free}
\end{eqnarray}
Here, $g_{m}=1+2 \tau_{m}$, $\tau_{m}$ is the integrated autocorrelation time 
with the parameter value $\lambda_m$ at temperature $T_{0}$, 
$N_{m}(E_{\lambda_1},\cdots,E_{\lambda_M};\xi_{1},\xi_{2})$
is the histogram of the $M$-dimensional energy distributions
at the parameter value $\lambda_m$ and the reaction
coordinate values $(\xi_{1},\xi_{2})$, which was obtained by
the Hamiltonian REM simulation,
$n_{m}$ is the total number of samples obtained at the
parameter value $\lambda_m$.
Note that this probability distribution is not normalized. 
Equations (\ref{prob}) and (\ref{dless_free}) are solved 
self-consistently by iteration. 
For biomolecular systems the quantity $g_m$ can safely be
set to be a constant in the reweighting formulas \cite{kbskr92},
and so we set $g_m =1$ throughout the analyses in the present
work.
Note also that these equations can be easily generalized 
to any reaction coordinates $(\xi_{1},\xi_{2},\cdots)$. 

>From the probability distribution 
$P_{T,\lambda}(\xi_{1},\xi_{2})$ in Eq.~(\ref{prob}), 
the expectation value of a physical quantity $A$ with 
any parameter value $\lambda$ at any temperature $T$
is given by 
\begin{eqnarray}
\left<A \right>_{T,\lambda} = 
\frac{\displaystyle \sum_{\xi_{1},\xi_{2}} A(\xi_{1},\xi_{2})
P_{T,\lambda}(\xi_{1},\xi_{2})}
{\displaystyle \sum_{\xi_{1},\xi_{2}} P_{T,\lambda}(\xi_{1},\xi_{2})}~.
\label{rew_ev}
\end{eqnarray}
We can also calculate the free energy (or, the potential of mean force) 
as a function of
the reaction coordinates $\xi_{1}$ and $\xi_{2}$ with 
any parameter value $\lambda$ at any temperature $T$ from
\begin{equation}
F_{T,\lambda}(\xi_{1},\xi_{2}) = 
-k_{\rm B}T{\rm ln}{P_{T,\lambda}(\xi_{1},\xi_{2})}~.
\label{def_free}
\end{equation}

By utilizing these equations, therefore, 
we can obtain various physical quantities 
from the Hamiltonian REM simulations with 
the original and non-original parameter values. 
We remark that although we wrote {\it any} $T$ in 
Eqs.~(\ref{prob}), (\ref{rew_ev}), and (\ref{def_free}) above,
the valid value $T$ is limited in the vicinity of $T_0$.
We also need the $T$-exchange process in order to have accurate
average quantities for a wide range of $T$ values.

  \section{Computational Details} 
  \label{det:sec}

In order to demonstrate the effectiveness of the present
Hamiltonian REM, namely, vWREM,  
in which we exchange pairs of the van der Waals radius parameter values, 
we applied the vWREM MD algorithm, which we refer to as the {\it vWREMD},
to the system of 
an alanine dipeptide in explicit water solvent and 
compared the results with those obtained by the conventional 
canonical MD simulation. 
The N-terminus and the C-terminus were blocked by 
the acetyl group and the N-methyl group, respectively. 
The number of water molecules was 67. 
The force field that we adopted was the AMBER parm96 parameter 
set \cite{amber96}, and 
the model for the water molecules was the TIP3P rigid-body model \cite{tip3p}. 
The vWREMD and canonical MD simulations were carried out with the 
symplectic integrator 
with rigid-body water molecules, 
in which the temperature was controlled by the 
Nos\'e-Poincar\'e thermostat \cite{bll99,nose01,oio07,oo07,oo08,o08}. 
The system was put in a cubic unit cell with the side length of
13.4 \AA, and we imposed the periodic boundary conditions. 
The electrostatic potential energy was calculated by the Ewald method, and 
we employed the minimum image convention for the Lennard-Jones 
potential energy. 
The time step was taken to be 0.5 fs. 

\begin{figure}
\includegraphics[width=6cm,keepaspectratio]{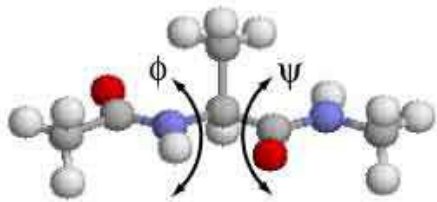}
\caption{
The common initial conformation of the alanine dipeptide.
The figure was created with RasMol \cite{rasmol}. 
}
\label{initial:fig}
\end{figure}

In the vWREMD simulation, we needed only four replicas ($M=4$). That is, 
we employed four different parameter values $\lambda_m$ $(m=1,\cdots,4)$. 
In Table~\ref{params:table} we list the values of these parameters. 
The original potential energy corresponds to the scale factor 
$\lambda_4=1.0$, and  
in the canonical MD simulation 
we employed this original parameter value of 1.0, 
The temperature of the system $T_0$ was set to be 300 K for
all the replicas in the vWREMD and in
the canonical MD simulation. 
Moreover, the initial conformations were also the same
for all the simulations, and 
the initial backbone dihedral angles $\phi$ and $\psi$ of 
the alanine dipeptide were set 
$(\phi,\psi)=(180^{\circ},180^{\circ})$ as shown in Fig.~\ref{initial:fig}. 
The total time of the MD simulations were 2.5 ns per replica
for the vWREMD simulation and 2.5 ns for the canonical simulation, 
including equilibration for 0.1 ns. 
The trajectory data were stored every 50 fs. 
The replica exchange was tried every 250 fs in the vWREMD simulation. 

  \section{Comparisons of the vWREMD simulation with the canonical MD simulation} 
  \label{comp:sec}

\begin{figure}
\includegraphics[width=13cm,keepaspectratio,angle=-90]{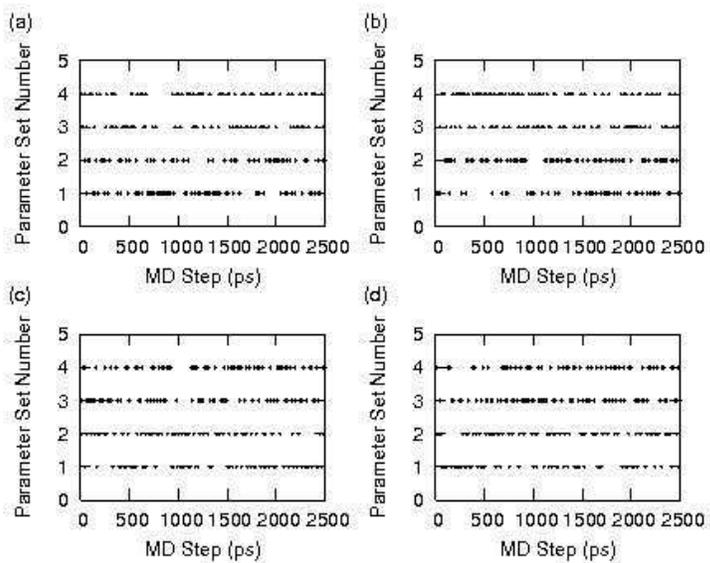}
\caption{
Time series of the label $m$ of the parameter $\lambda_m$
($m=1, 2, 3, 4$) in (a) Replica 1, 
(b) Replica 2, (c) Replica 3, and (d) Replica 4
during the vWREMD simulation. 
}
\label{time_param:fig}
\end{figure}

We first examine whether the exchanges of pairs of the parameter values 
were realized sufficiently in our vWREMD simulation. 
In Table~\ref{accept:table} 
we list the acceptance ratios of replica exchange of the parameter values 
in the vWREMD simulation. 
These acceptance ratios are large enough ($>$ 40 $\%$) except for
the pair of $\lambda_4$ and $\lambda_1$. 
Because the difference of these parameter values is much larger
than those of other pairs, this low acceptance ratio is expected,
and it does not affect the REM performance. 
Figure~\ref{time_param:fig} shows the time series of the parameter 
set number $m$ in $\lambda_m$ which each replica visited. 
This figure shows that random walks in the parameter space were realized 
in the vWREMD simulation. 

\begin{figure}
\includegraphics[width=13cm,keepaspectratio,angle=-90]{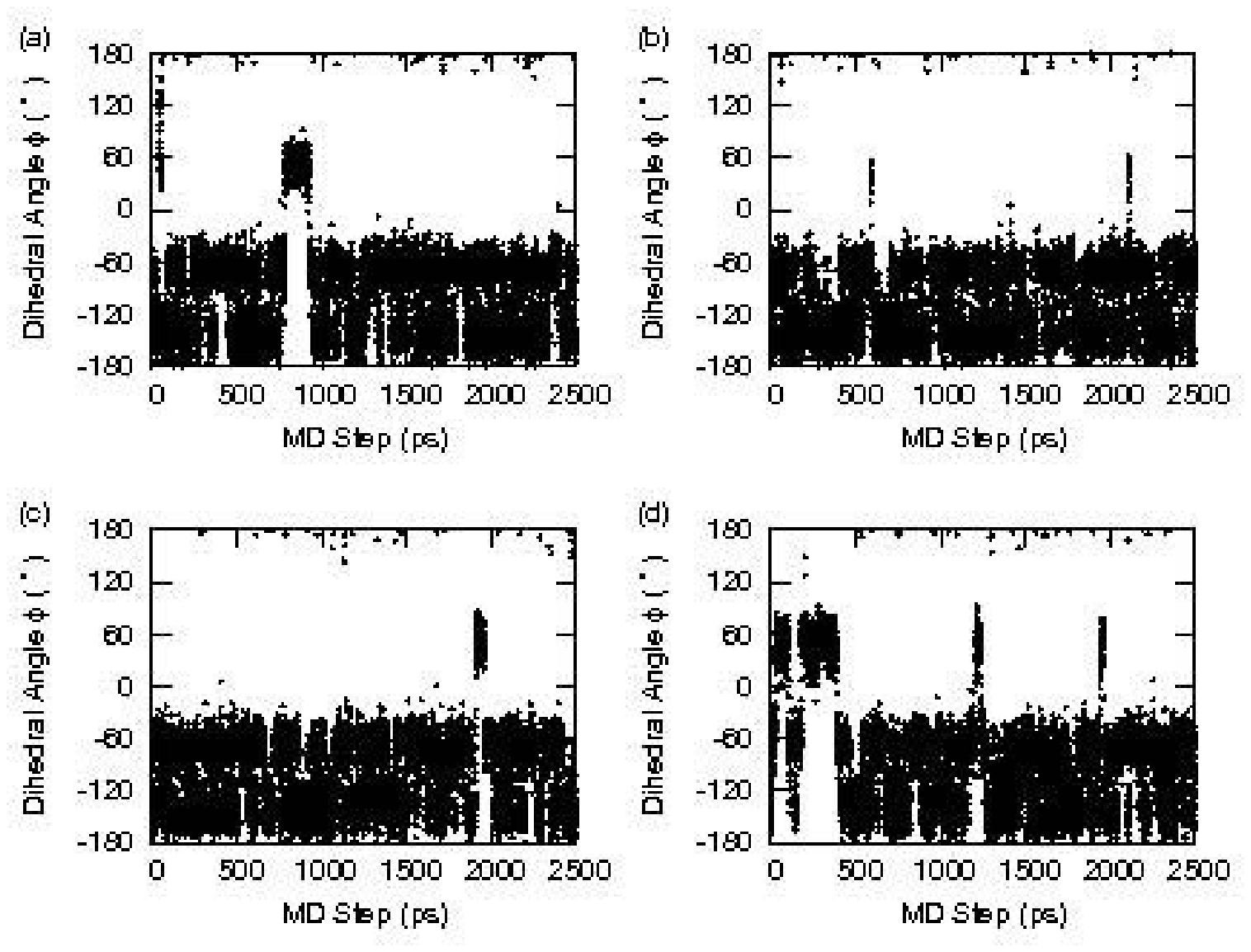}
\caption{
Time series of the dihedral angles $\phi$ in (a) Replica 1, 
(b) Replica 2, (c) Replica 3, and (d) Replica 4 
during the vWREMD simulation.
}
\label{time_phi_rep:fig}
\end{figure}

\begin{figure}
\includegraphics[width=13cm,keepaspectratio,angle=-90]{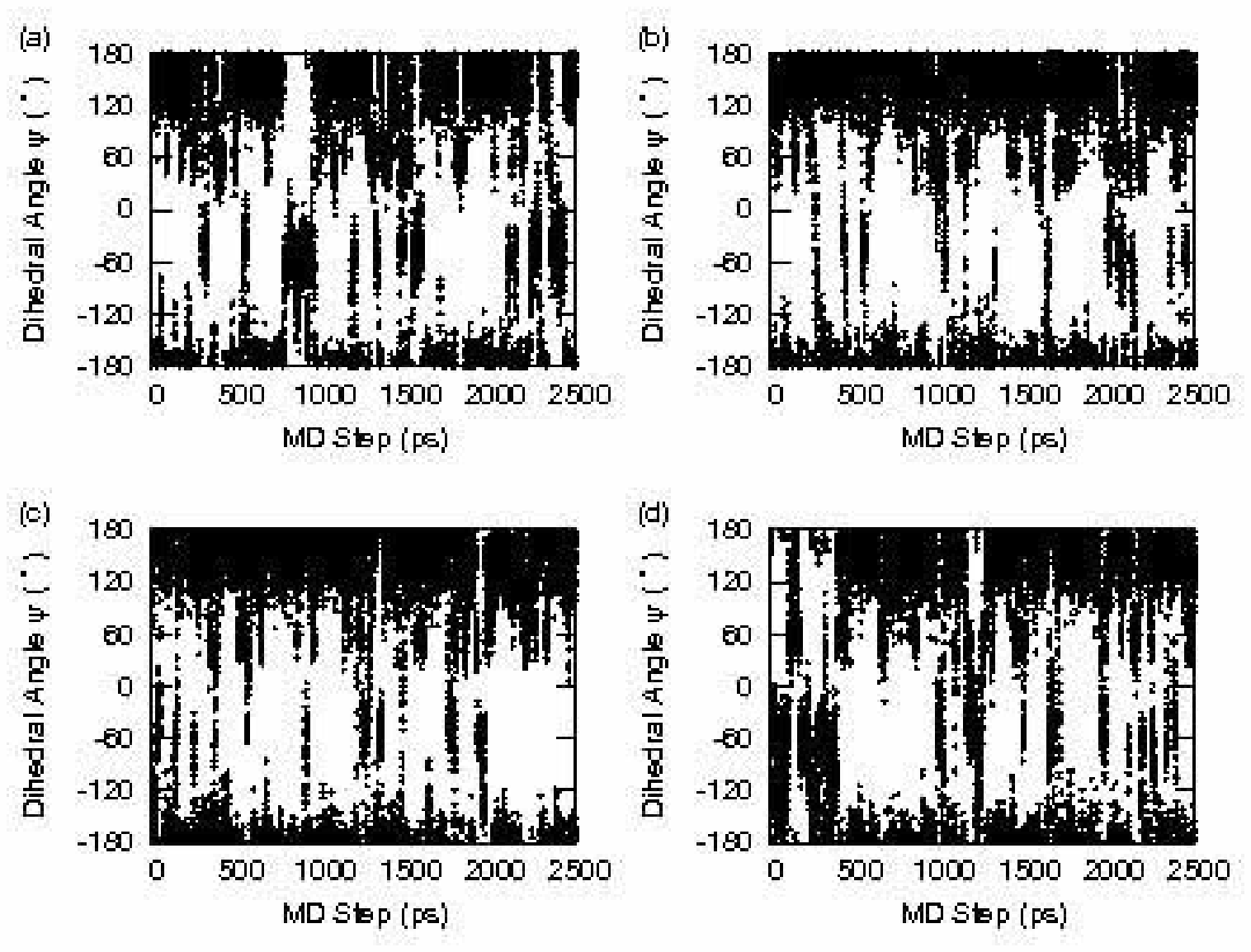}
\caption{
Time series of the dihedral angles $\psi$ in (a) Replica 1, 
(b) Replica 2, (c) Replica 3, and (d) Replica 4 
during the vWREMD simulation.
}
\label{time_psi_rep:fig}
\end{figure}

\begin{figure}
\includegraphics[width=6cm,keepaspectratio]{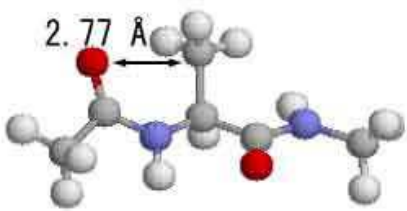}
\caption{
The snapshot of the alanine dipeptide at 2,469 ps in Replica 3 
in the vWREMD simulation. 
The distance between the ${\rm C}_{\beta}$ atom of the alanine and  
the O atom of the acetyl group is presented. 
The figure was created with RasMol \cite{rasmol}. 
}
\label{restricted:fig}
\end{figure}

Figures~\ref{time_phi_rep:fig} and \ref{time_psi_rep:fig} show the 
time series of 
the backbone-dihedral angles $\phi$ and $\psi$ for each replica. 
Figure~\ref{time_phi_rep:fig} indicates that it is difficult to sample 
the range 
of the dihedral angle $\phi$ between $90^{\circ}$ and $180^{\circ}$. 
This is because the steric restriction between the ${\rm C}_{\beta}$ atom 
of the alanine and  
the O atom of the acetyl group prevent the rotation in this range. 
For instance, 
Fig.~\ref{restricted:fig} is the snapshot of the alanine dipeptide 
which corresponds to the structure at 2,469 ps in Replica 3.
In this structure the angle $\phi$ is $148.8^{\circ}$, and 
the distance between the ${\rm C}_{\beta}$ atom of the alanine and  
the O atom of the acetyl group is 2.77 \AA. 
For such a short distance, the two atoms collide with each other. 
Therefore, the sampling among the range 
of $90^{\circ} < \phi < 180^{\circ}$ in 
the vWREMD simulation was rare,  
although the scale factor $\lambda_m$ for the van der Waals
radii was lessened.
In order to sample this range frequently, 
it is necessary to employ a much smaller scale factor than 
the present case.
However, conformations among this range have quite high potential energy 
due to the collisions between the ${\rm C}_{\beta}$ atom of the alanine and  
the O atom of the acetyl group, 
and it is not so important to sample this range at room temperature. 
On the other hand, the vWREMD simulation 
realized effective samplings 
with respect to $\psi$ in all the replicas as shown in 
Fig.~\ref{time_psi_rep:fig}. 
This is because the van der Waals radius of the H atom of the 
N-methyl group that have 
the covalent bond with the N atom of the N-methyl group is small, and 
steric restrictions between the ${\rm C}_{\beta}$ atom of the alanine 
and the H atom 
are less than those between the ${\rm C}_{\beta}$ atom of the alanine and  
the O atom of the acetyl group. 

\begin{figure}
\includegraphics[width=13cm,keepaspectratio,angle=-90]{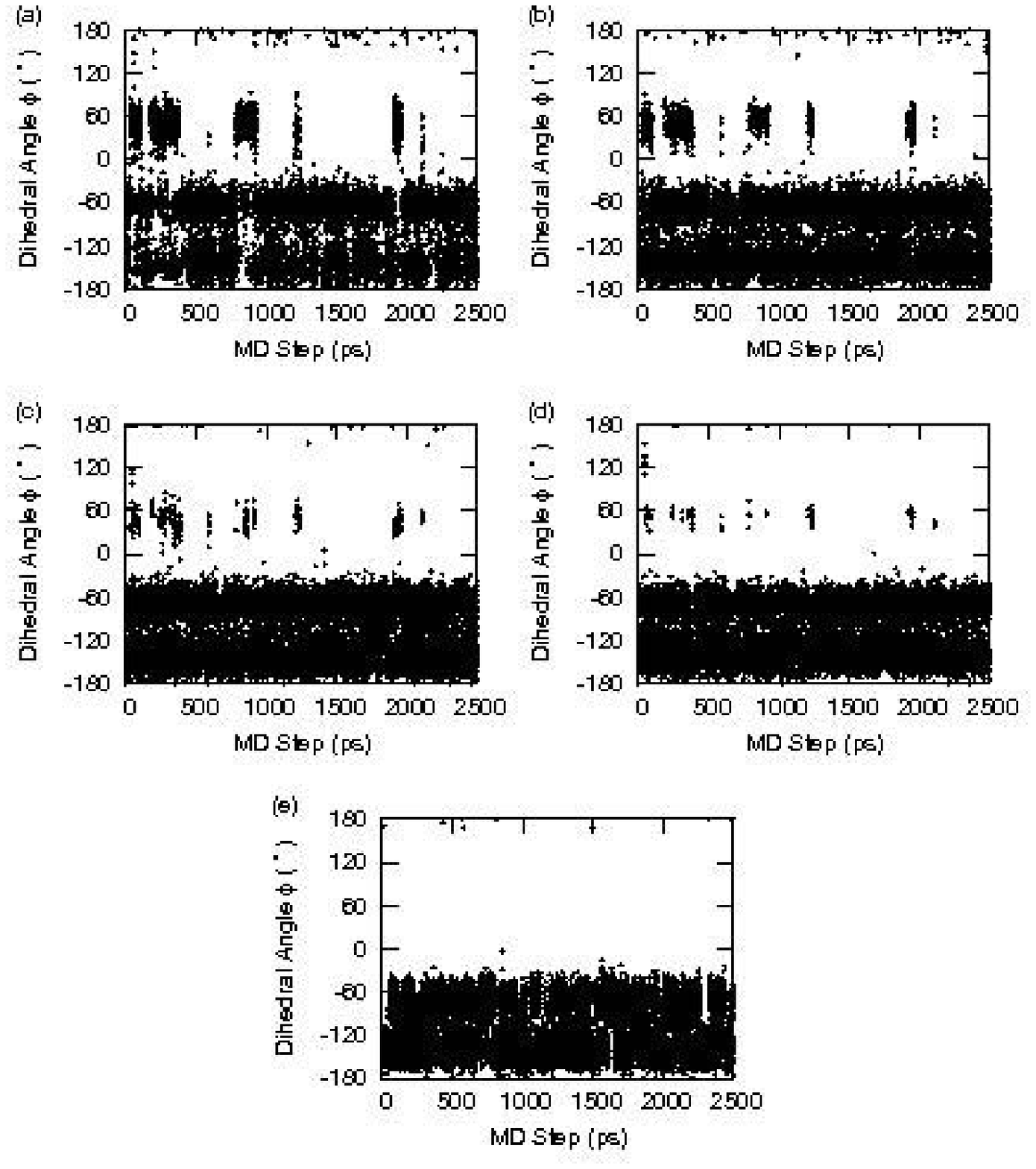}
\caption{
Time series of the dihedral angle $\phi$ with the
parameter values (a) $\lambda_1$, 
(b) $\lambda_2$, (c) $\lambda_3$, and (d) $\lambda_4$ 
during the vWREMD simulation. 
(e) Time series of the dihedral angle $\phi$ during
the conventional canonical MD simulation. 
}
\label{time_phi_prm:fig}
\end{figure}

\begin{figure}
\includegraphics[width=13cm,keepaspectratio,angle=-90]{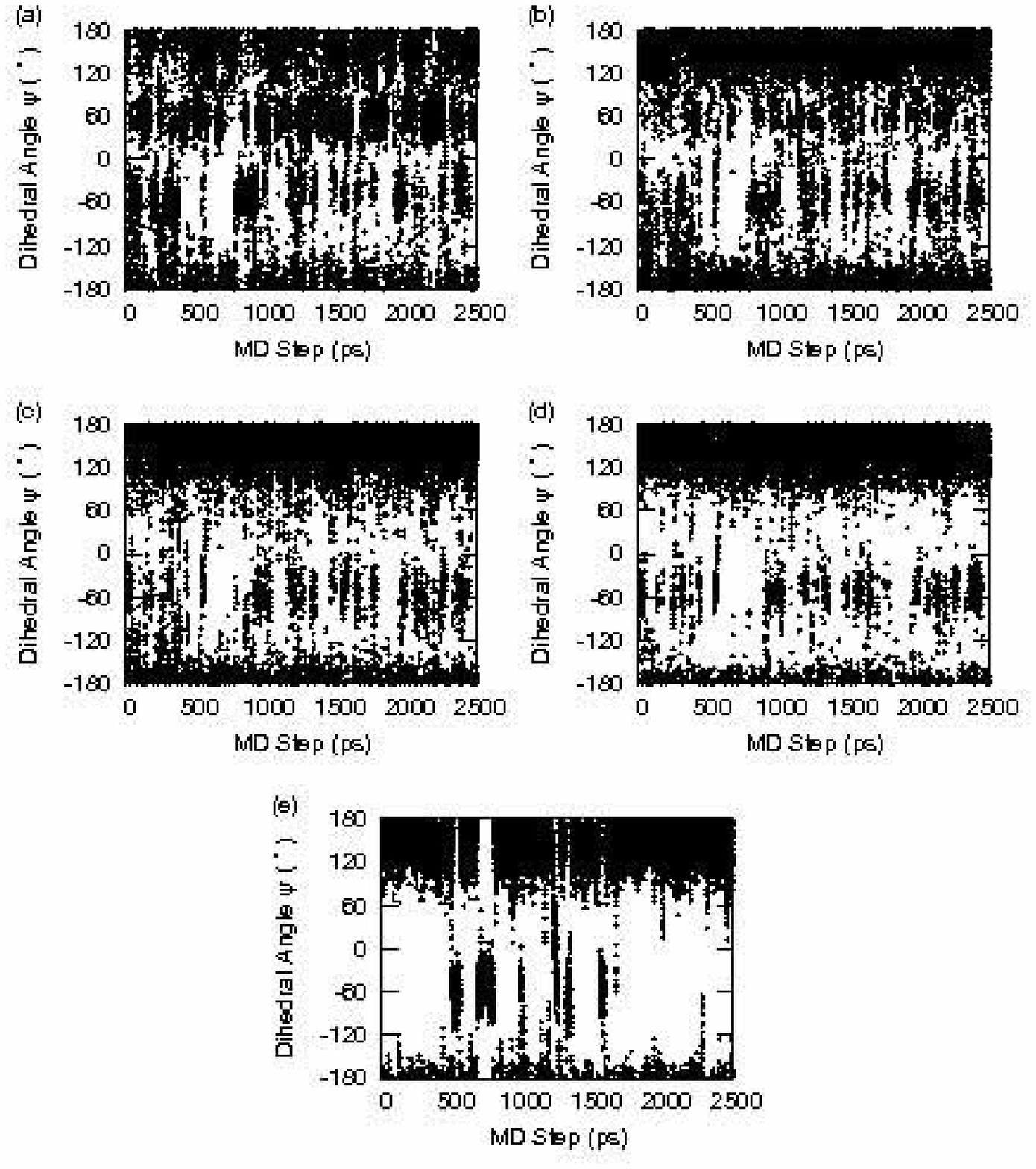}
\caption{
Time series of the dihedral angle $\psi$ with the
parameter values (a) $\lambda_1$, 
(b) $\lambda_2$, (c) $\lambda_3$, and (d) $\lambda_4$ 
during the vWREMD simulation. 
(e) Time series of the dihedral angle $\psi$ during
the conventional canonical MD simulation. 
}
\label{time_psi_prm:fig}
\end{figure}

Figures~\ref{time_phi_prm:fig} and \ref{time_psi_prm:fig} show the 
time series of 
the backbone-dihedral angles $\phi$ and $\psi$ for each parameter value. 
For comparisons, we also show those from the conventional MD simulation
with the original parameter value of $\lambda=1$.
These figures show that the smaller the scale factor $\lambda_m$
of the van der Waals radii is, 
the more efficient the sampling in the backbone-dihedral-angle space 
is. 
This is because the steric restrictions are reduced by lessening 
the scale factor and the 
energy barriers caused by the steric restrictions are decreased. 
Comparing Fig.~\ref{time_phi_prm:fig}(d) with 
Fig.~\ref{time_phi_prm:fig}(e) and 
Fig.~\ref{time_psi_prm:fig}(d) with 
Fig.~\ref{time_psi_prm:fig}(e), 
moreover, the vWREMD with 
the original parameter value $\lambda_4$ 
sampled the dihedral-angle space more effectively than the usual 
canonical MD simulation 
and did not get trapped in the local-minimum free-energy states. 
In other words, the canonical MD simulation with the original 
parameter value 
could not overcome energy barriers caused by the steric restrictions and 
got trapped in the local-minimum free-energy states. 
Therefore, effective samplings in the dihedral-angle space cannot be realized 
in the usual canonical MD simulations. 

\begin{figure}
\includegraphics[width=13cm,keepaspectratio]{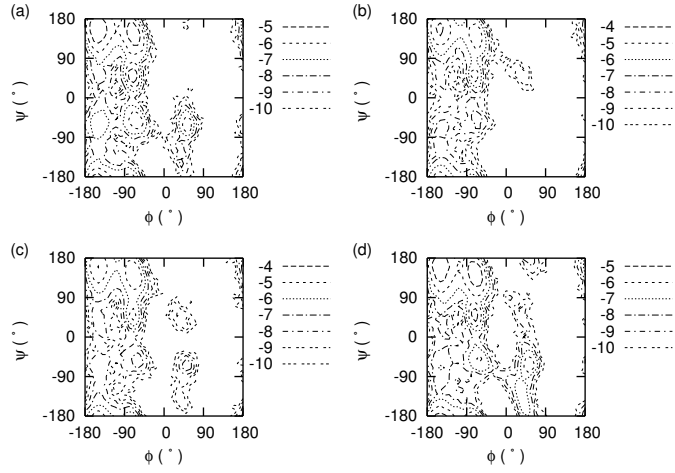}
\caption{
The logarithm of the probability distribution ${\rm ln} P(\phi,\psi)$  
with respect to the dihedral angles $\phi$ and $\psi$ 
for (a) Replica 1, (b) Replica 2, (c) Replica 3, and (d) Replica 4
in the vWREMD simulation. 
}
\label{rmch_rep:fig}
\end{figure}

\begin{figure}
\includegraphics[width=13cm,keepaspectratio]{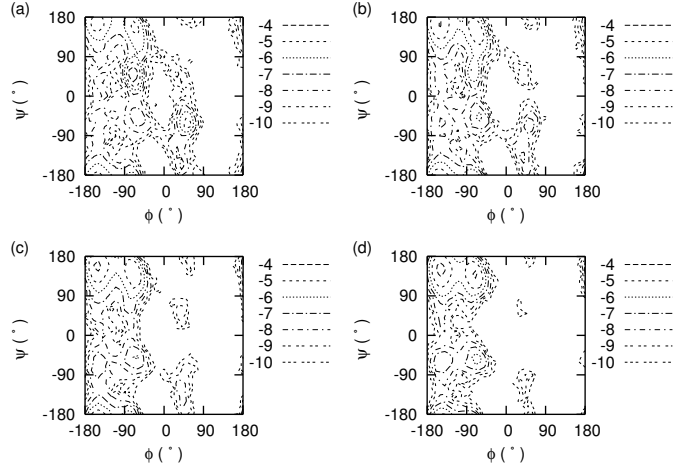}
\caption{
The logarithm of the probability distribution ${\rm ln} P(\phi,\psi)$ 
with respect to the dihedral angles $\phi$ and $\psi$ 
with parameter values (a) $\lambda_1$, (b) $\lambda_2$, (c) $\lambda_3$, 
and (d) $\lambda_4$ 
in the vWREMD simulation. 
}
\label{rmch_prm:fig}
\end{figure}

\begin{figure}
\includegraphics[width=6cm,keepaspectratio]{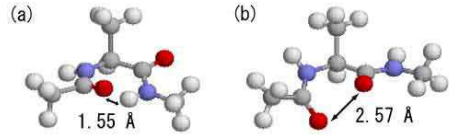}
\caption{
(a) The snapshot of the alanine dipeptide at 895.55 ps in Replica 1 
in the vWREMD simulation. 
The structure has backbone-dihedral angles 
of $(\phi,\psi)=(15.5^{\circ},5.2^{\circ})$. 
The distance between the O atom of the acetyl group and 
the H atom of N-methyl group is given.
(b) The snapshot of the alanine dipeptide at 2,465.65 ps in Replica 1 
in the vWREMD simulation. 
The structure has backbone-dihedral angles 
of $(\phi,\psi)=(-27.7^{\circ},174.1^{\circ})$. 
The distance between the O atom of the acetyl group and 
the O atom of the alanine is given.
The figures were created with RasMol \cite{rasmol}. 
}
\label{restricted1:fig}
\end{figure}

The logarithm of the probability distributions, ${\rm ln} P(\phi,\psi)$,  
with respect to dihedral angles $\phi$ and $\psi$ were obtained from 
the vWREMD simulations. 
Figures~\ref{rmch_rep:fig} and \ref{rmch_prm:fig} show 
${\rm ln} P(\phi,\psi)$ 
for each replica and those for each parameter value, respectively. 
From Fig.~\ref{rmch_rep:fig} we see that there are two regions in the vicinites of 
$(\phi,\psi)=(0^{\circ},0^{\circ})$ and $(\phi,\psi)=(0^{\circ},180^{\circ})$ 
in which the samplings were rare except the range of $90^{\circ} < \phi < 180^{\circ}$. 
This is because in the vicinity of 
$(\phi,\psi)=(0^{\circ},0^{\circ})$, the O atom of the acetyl group and 
the H atom of N-methyl group collide with each other as in a structure 
shown in Fig.~\ref{restricted1:fig}(a). 
In the region of the neighborhood of $(\phi,\psi)=(0^{\circ},180^{\circ})$, 
the O atom of the acetyl group and the O atom of the alanine also collide 
as in a structure shown in Fig.~\ref{restricted1:fig}(b). 
It is obvious that steric restrictions in these regions were reduced 
and that the energy barriers were decreased by decreasing the scale factor 
as shown in Fig.~\ref{rmch_prm:fig}. 
However, the effects of reducing energy barriers 
around $(\phi,\psi)=(0^{\circ},180^{\circ})$ are less than those around 
$(\phi,\psi)=(0^{\circ},0^{\circ})$. 
This is because the O atom of the acetyl group and the O atom of the alanine 
have negative charges, and repulsive forces caused by these charges act on both atoms. 
Therefore, by reducing the atomic charges as well as the
van der Waals radii, more effective sampling 
may be realized in the conformational space. 

\begin{figure}
\includegraphics[width=13cm,keepaspectratio]{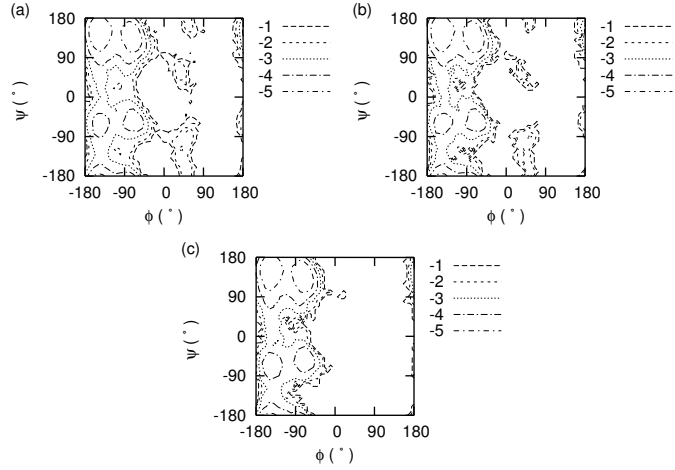}
\caption{
The free-energy landscapes at $T_0=300$ K with respect to 
the backbone-dihedral angles $\phi$ and $\psi$ 
with the original parameter value $\lambda = 1$. 
These were obtained from (a) the reweighting techniques
applied to the results of the vWREMD simulation, 
(b) the raw histogram with the original parameter value
$\lambda_4$ 
in the vWREMD simulation, 
and (c) the raw histogram in the canonical MD simulation. 
}
\label{rmch_cr:fig}
\end{figure}

Figure~\ref{rmch_cr:fig} shows the free-energy landscapes 
at $T_0=300$ K 
with respect to the backbone-dihedral angles $\phi$ and $\psi$ 
with the original parameter value ($\lambda = 1$). 
The free-energy landscape in Fig.~\ref{rmch_cr:fig}(a) was calculated from 
Eq.~(\ref{def_free}) by 
the reweighting techniques in Eqs.~(\ref{prob}) and (\ref{dless_free}).
The free-energy landscapes in Fig.~\ref{rmch_cr:fig}(b) and 
Fig.~\ref{rmch_cr:fig}(c) were obtained from 
the raw histogram with the original parameter value $\lambda_4$
in the vWREMD simulation 
and the raw histogram in the canonical MD simulation, respectively. 
The free-energy landscape of the conventional canonical MD simulation 
is inaccurate due to insufficient sampling in the backbone-dihedral-angle space 
as shown in Figs.~\ref{time_phi_prm:fig}(e) and \ref{time_psi_prm:fig}(e). 
The free-energy landscape obtained by the reweighting techniques 
in Fig.~\ref{rmch_cr:fig}(a) shows a better statistics even at 
free-energy barriers
among the local-minimum states in comparison with that 
obtained from the raw histogram 
in Fig.~\ref{rmch_cr:fig}(b). 
This is because the information of all the other parameter values 
can be reflected by the reweighting techniques. 
Therefore, more accurate free-energy landscape can be obtained by 
the reweighting techniques. 

  \section{Conclusions} 
  \label{conc:sec}

In this article, we introduced a new type of Hamiltonian REM, the
van der Waals REM (vWREM) in which
the scale factor of the van der Waals radii of the solute atoms
is exchanged only in the Lennard-Jones interactions among themselves.
The steric hindrance due to the Lennard-Jones repulsions
can be reduced by this method. 
Accordingly, the vWREM simulation can realize effective sampling 
in the backbone-dihedral-angle space without getting trapped in 
local-minimum free-energy states 
in comparison with the conventional canonical MD simulation. 
Employing the reweighting techniques, furthermore, 
we can obtain accurate free-energy landscape. 

Although we considered only exchanges of the scale factor of
the van der Waals radii in this article, 
this idea can be extended to other parameters. For example, 
the scale factor of partial charges of solute atoms 
so that we can also realize even more efficient sampling 
in the conformational space. 
The generalization of the formalism in Sec.~\ref{method:sec} to 
other parameters is straightforward 
including the reweighting techniques. 
Moreover, these formalisms are independent of the degrees of freedom 
of solvent molecules. 
Therefore, these algorithms can be easily applied to large biomolecular
systems. 

\section*{ACKNOWLEDGMENTS}

The computations were performed on the computers at 
the  esearch Center for Computational Science, Institute for Molecular Science. 
This work was supported, in part, by Grants-in-Aid
for Scientific Research on Innovative Areas 
(``Fluctuations and Biological Functions'' )
and for the Next-Generation Super Computing Project, Nanoscience Program
from the Ministry of Education, Culture, Sports, Science and Technology (MEXT), Japan.



%
\begin{table}
\caption{The values of the scaling factor, $\lambda_m$ ($m=1, \cdots, 4$), 
of the van der Waals radii in the vWREMD simulation.
}
\label{params:table}
\vspace{0.5cm}
\begin{tabular}{c|c}
\hline \hline
Parameter $\lambda_m$ & Parameter value \\
\hline
$\lambda_1$ & 0.85 \\
$\lambda_2$ & 0.90 \\
$\lambda_3$ & 0.95 \\
$\lambda_4$ & 1.00 \\
\hline \hline
\end{tabular}

\vspace{0.5cm}
\end{table}

\begin{table}
\caption{Acceptance ratios of replica exchange between
pairs of the neighboring parameter values. 
}
\label{accept:table}
\vspace{0.5cm}
\begin{tabular}{c|c}
\hline \hline
Pair of parameter values & Acceptance ratio \\
\hline
$\lambda_1$ $\leftrightarrow$ $\lambda_2$ & 0.482 \\
$\lambda_2$ $\leftrightarrow$ $\lambda_3$ & 0.540 \\
$\lambda_3$ $\leftrightarrow$ $\lambda_4$ & 0.442 \\
$\lambda_4$ $\leftrightarrow$ $\lambda_1$ & 0.055 \\
\hline \hline
\end{tabular}

\vspace{0.5cm}
\end{table}


%
%
%
%
%
%
%
%
%
%
%
%
%
%
%
%
%
%
%
%
%
\end{document}